\documentclass{iopart}
\usepackage{iopams}
\usepackage{graphicx}
\usepackage{color}

\begin{document}
\title{Beyond the Second Generation of Laser-Interferometric Gravitational Wave 
Observatories}
\author{
S\,Hild
}

\ead{stefan.hild@glasgow.ac.uk}
\vskip 1mm
\address{SUPA, School of Physics and Astronomy, The University of Glasgow, Glasgow, G12\,8QQ, UK}

\begin{abstract}
This article gives an overview of potential upgrades of
second generation gravitational wave detectors and the required
key technologies to improve the limiting noise sources. In addition
the baseline design of the Einstein Telescope, a European third generation 
gravitational wave observatory, is briefly discussed. 
\end{abstract}


\section{Introduction}
\label{sec:intro}

Over the last two decades major advances have been accomplished in  high-precision interferometry, 
targeting the direct observation of gravitational wave signals from astrophysical sources.
A network  of kilometre-scale laser-interferometric gravitational wave detectors (LIGO \cite{ligo}, 
Virgo \cite{virgo}, 
Tama \cite{tama} and GEO\,600 \cite{geo}) has been constructed and
has 
collected years worth of data from coincident observation at unprecedented sensitivity
\cite{crab, stochastic, magnetars,  GRB}. 

\begin{figure}[htbp]
\centering
\includegraphics[width=0.8\textwidth]{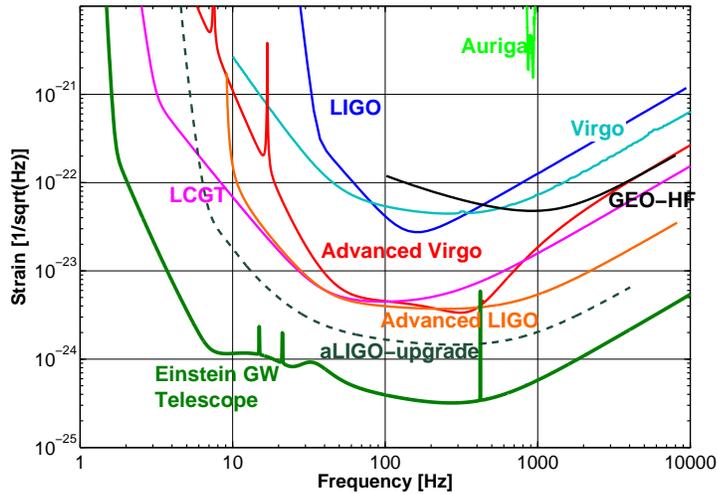}
\caption{Sensitivity curves for past, present and future GW observatories. The first generation 
of GW detectors is shown in blueish colours, second generation in reddish colours and
third generation in green. All traces originate from \cite{GWIC-RM}, apart from the traces 
labeled ALIGO-upgrade (which is the potential sensitivity of an Advanced LIGO upgrade)
 \cite{Rana} and the Einstein Telescope \cite{ET-D}. }\label{fig:GWIC}
\end{figure}

Currently major programs 
are underway to upgrade these instruments (see \cite{GL_amaldi} in this journal for 
details) and establish 
the so-called second generation of gravitational wave detectors (Advanced LIGO \cite{aligo}, 
GEO-HF \cite{Willke06}, LCGT \cite{lcgt06} and Advanced Virgo \cite{avirgo}).  On reaching 
their target sensitivities in the second half of this decade, these advanced detectors are expected
to ensure the first direct detection of gravitational waves \cite{rates}.
While this will mark the beginning of gravitational wave astronomy,
only upgrades to the second generation instruments \cite{LSC-WP} and
subsequently construction of the third generation instruments, such as the
proposed Einstein Telescope \cite{et_punturo2010, ET}, will allow us to observe
high-SNR gravitational wave signals from astrophysical
sources on a regular basis. Figure \ref{fig:GWIC} shows the
design sensitivity curves of various gravitational wave
detectors, with blueish, reddish and greenish colours indicating first, second
and third  generation instruments, respectively.
For a summary of what science in terms of
astro-physics, cosmology and fundamental physics will come
into our reach with a third generation instrument, such as
the Einstein Telescope (ET), please see \cite{ET_DS,
Sathya_amaldi}.

This article gives an overview of the techniques required to successfully advance 
beyond the second generation of laser interferometric gravitational
observatories. In addition some experimental challenges, especially
associated with the fundamental noise limitations, are
discussed.

\section{Paths for the reduction of fundamental noise sources}
\label{sec:facility}

\begin{figure}[htbp]
\centering
\includegraphics[width=0.49\textwidth]{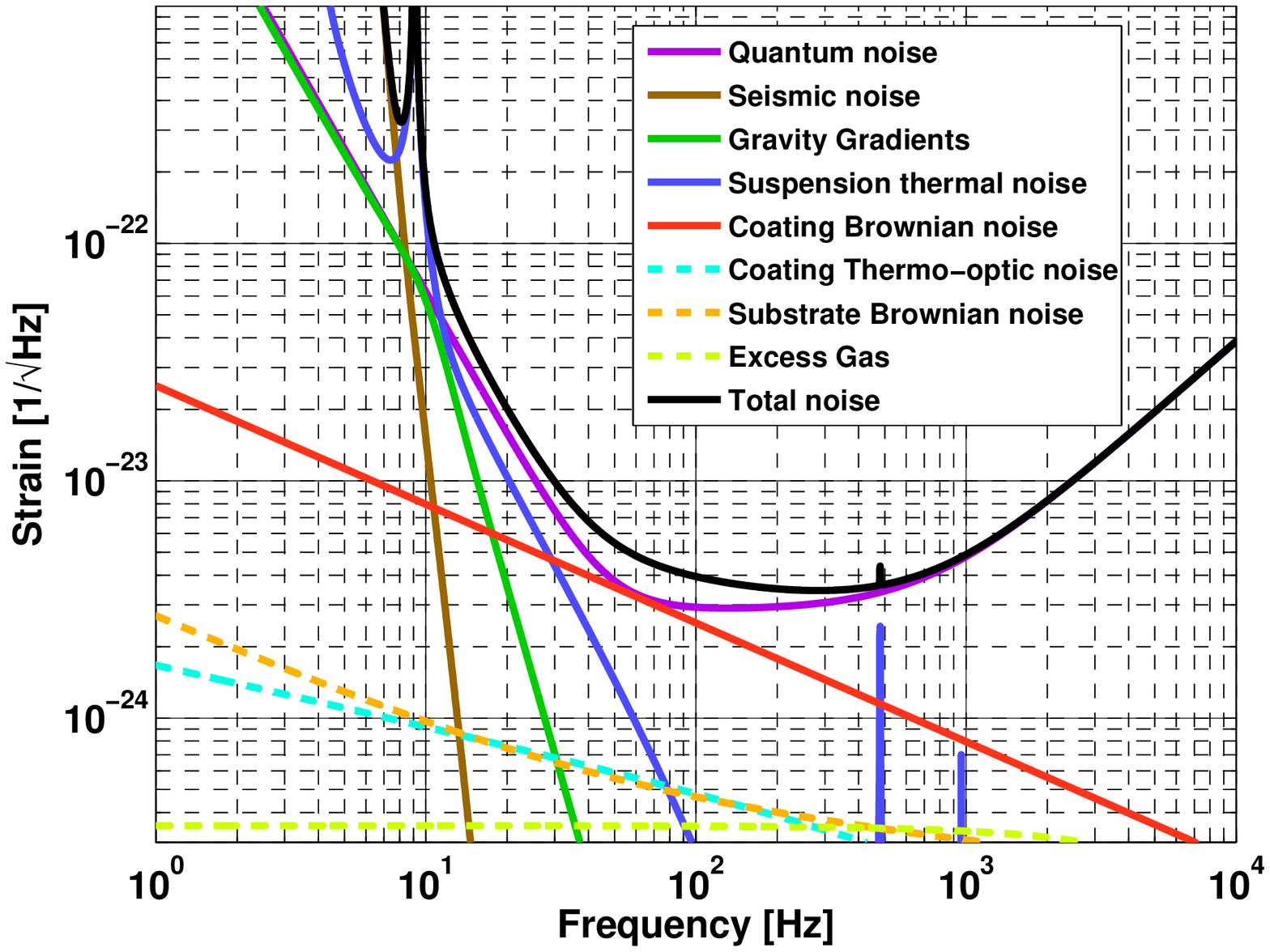}
\includegraphics[width=0.49\textwidth]{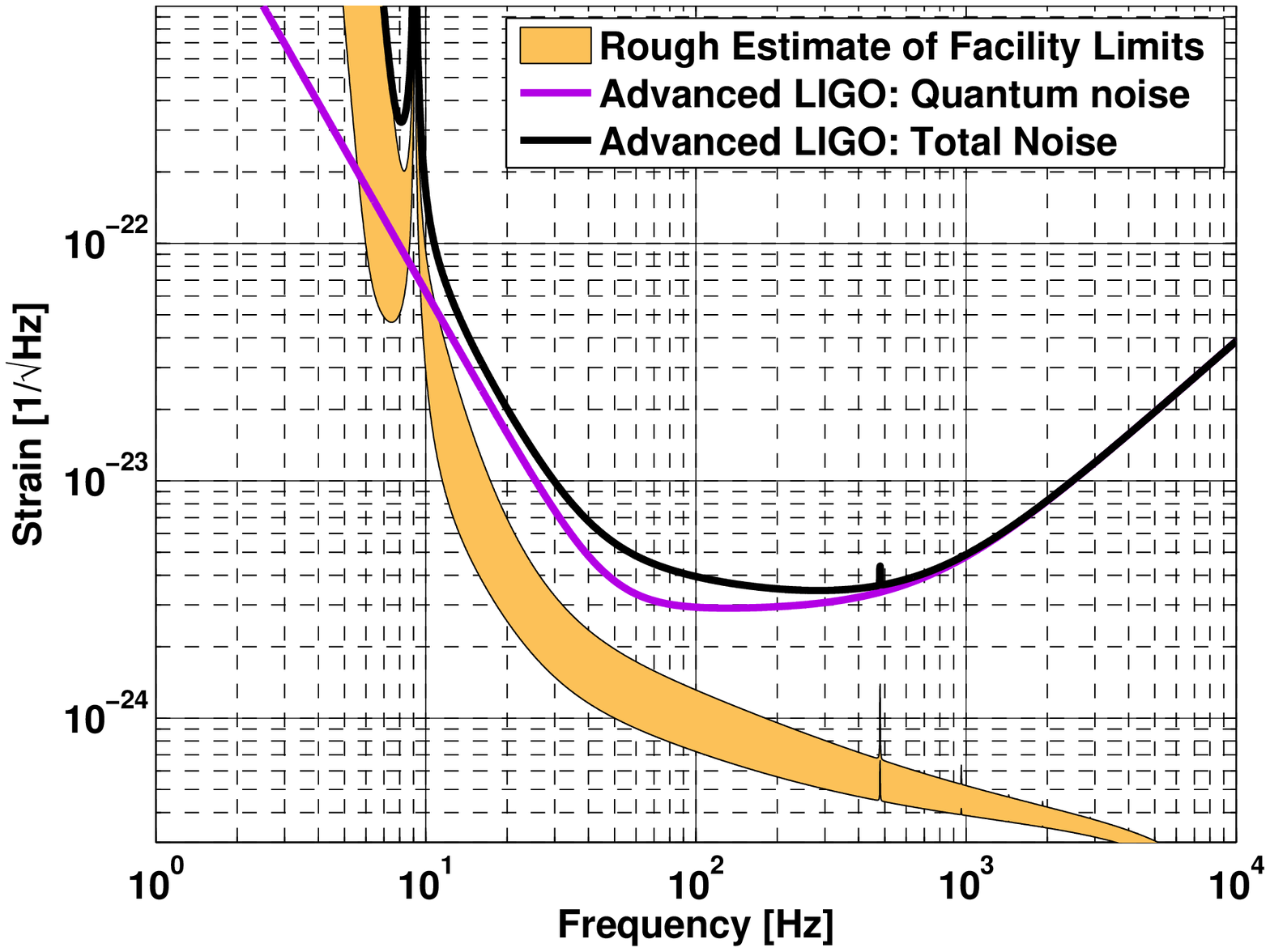}
\caption{LEFT: Noise budget of Advanced LIGO. This plot was produced using the GWINC 
\cite{website_GWINC} and represents the Advanced LIGO broadband configuration 
described in \cite{T070247-01}. RIGHT: Illustrative examples of potential sensitivity limits for 
Advanced LIGO upgrades. The upper boundary of the orange area is given
by seismic, gravity gradient and residual gas noise equal to the Advanced LIGO 
baseline design and coating 
and suspension thermal noise being improved by a factor 2 each. In contrast the 
lower boundary is calculated assuming a coating noise improvement of 
a factor 4, a suspension thermal noise reduction of a factor 5, a gravity gradient subtraction
of a factor 10 and a seismic noise level reduced by a factor 100. Please note that 
quantum noise is not included in the orange region.}\label{fig:ALIGO}
\end{figure}

In order to understand how we can proceed beyond the second
generation of laser-interferometric gravitational wave (GW)
detectors, we have to understand by which noise sources  instruments like
Advanced LIGO will be limited. The left plot of Figure~\ref{fig:ALIGO} shows 
the contributions of fundamental noise sources (coloured traces) to the 
advanced LIGO sensitivity (black trace) \cite{T070247-01,website_GWINC}. 
Here the term fundamental noise source refers to instrument-inherent
 noise sources, characterised by the actual technical implementation of the GW detector
 (such as the thermal noise of the mirror coatings or 
the seismic noise on the test masses). In contrast to fundamental noise sources,
the term technical noise source is applied to noise sources, such
as beam jitter or laser frequency noise, which 
 can in principle be reduced by implementing an improved performance
of the corresponding subsystem. The advanced LIGO design
sensitivity is limited over nearly the entire detection
band, i.e., for all frequencies above 12\,Hz,
by quantum noise \cite{Buonanno02} which consists of photon shot noise at high
frequencies and photon radiation pressure noise at low
frequencies. In the range from about 50--100\,Hz coating
Brownian noise \cite{Harry06}  is close to limiting the Advanced LIGO
sensitivity, while at the low-frequency end of the detection
band the limit is a mixture of thermal noise in
the fused silica suspension fibres \cite{Husman}, gravity gradient noise
\cite{Saulson84, Beccaria98, Hughes98} and seismic noise.
The remaining three noise traces included in the left hand
plot  of Figure \ref{fig:ALIGO}, Brownian thermal noise of the mirror 
substrates, coating thermo-optic noise \cite{Evans08} and excess noise
from residual gas inside the vacuum systems \cite{gasnoise} only play a secondary 
role for the advanced LIGO baseline design.  

In general, for each fundamental noise source there are
several ways to further reduce it and by that improve the
sensitivity beyond the advanced LIGO target sensitivity.
These potential improvements vary extremely in terms of 
implementation cost and required hardware effort.

\begin{itemize}
  \item{\textbf{Quantum noise:}  
There are various ways to decrease the quantum noise, at least in a specific 
frequency region. Increasing the light power inside the interferometer arms 
reduces the shot noise level, but at the same time increases the radiation
pressure noise. Signal recycling \cite{Meers88} allows the quantum 
noise contribution to be shaped to optimise the overall detector response. The signal recycling
bandwidth and the signal recycling tuning (i.e.~the frequency of maximum 
sensitivity) can be adjusted by means of the reflectivity
and microscopic position of the signal recycling mirror \cite{Hild07a}. Moreover,
the injection of squeezed light states \cite{Caves81} allows us to further manipulate the 
quantum noise level \cite{LSC11} (see left plot of Figure \ref{fig:squeez-level}).
 The techniques
mentioned so far require only rather small hardware changes.
Other more hardware intensive ways to further reduce quantum
noise include the application of heavier test masses, yielding a reduced susceptibility 
to quantum radiation pressure noise, the injection of frequency dependent squeezed
light \cite{Kimble02} and a multitude of other quantum-non-demolition techniques, 
such as optical bar \cite{brag96, Khalili02, Rehbein} and speed-meter 
\cite{Chen03} configurations. Please note that the latter
techniques might require a close-to-complete reorganisation of the
interferometer configuration inside the vacuum facilities (see Figure 
\ref{fig:speedmeter}).
It is also worth mentioning that most of these techniques are
not mutually exclusive, but any GW detector beyond the second
generation is likely to employ a `cocktail' of the above
mentioned techniques
 }

\begin{figure}[htbp]
\centering
\includegraphics[width=0.49\textwidth]{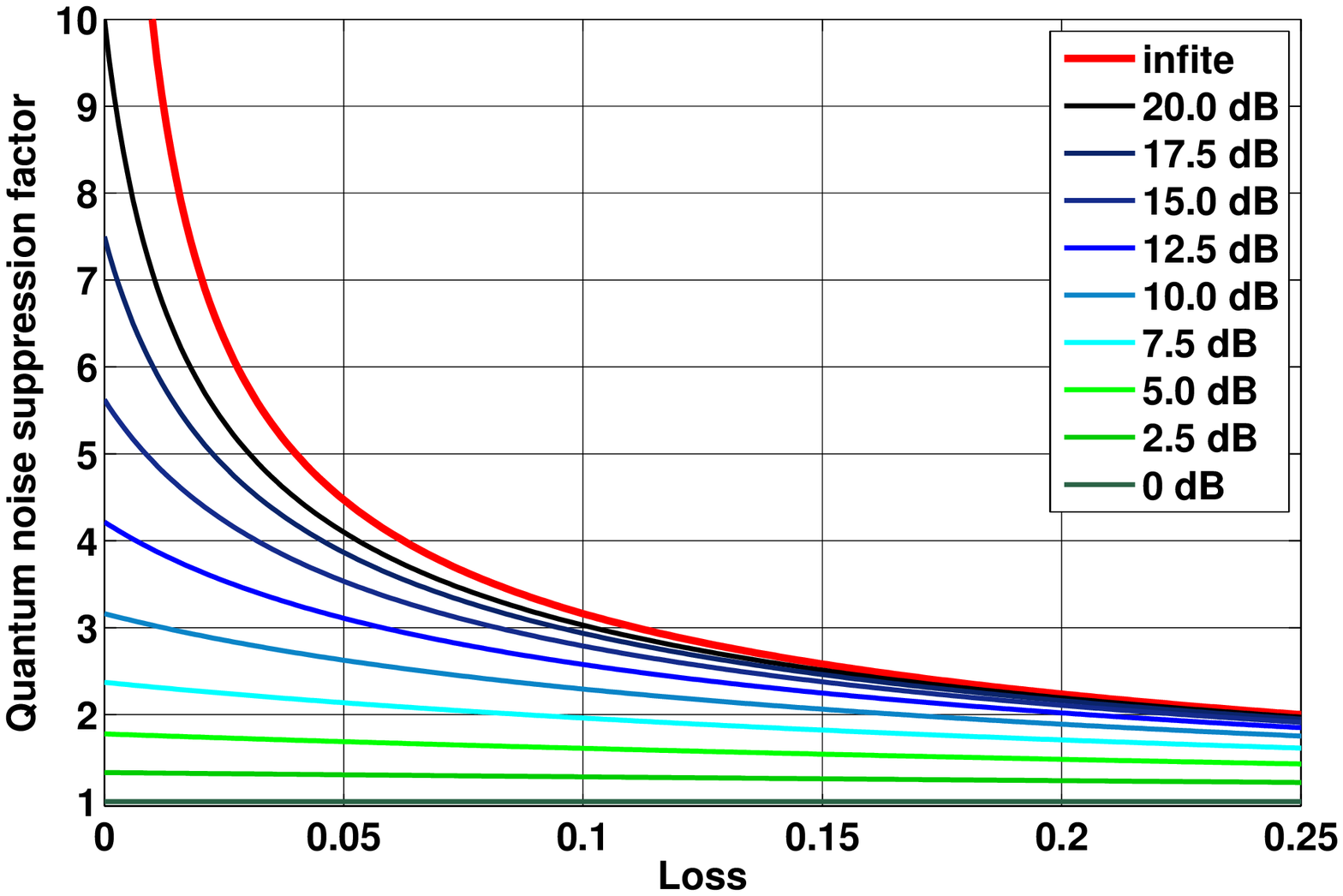}
\includegraphics[width=0.49\textwidth]{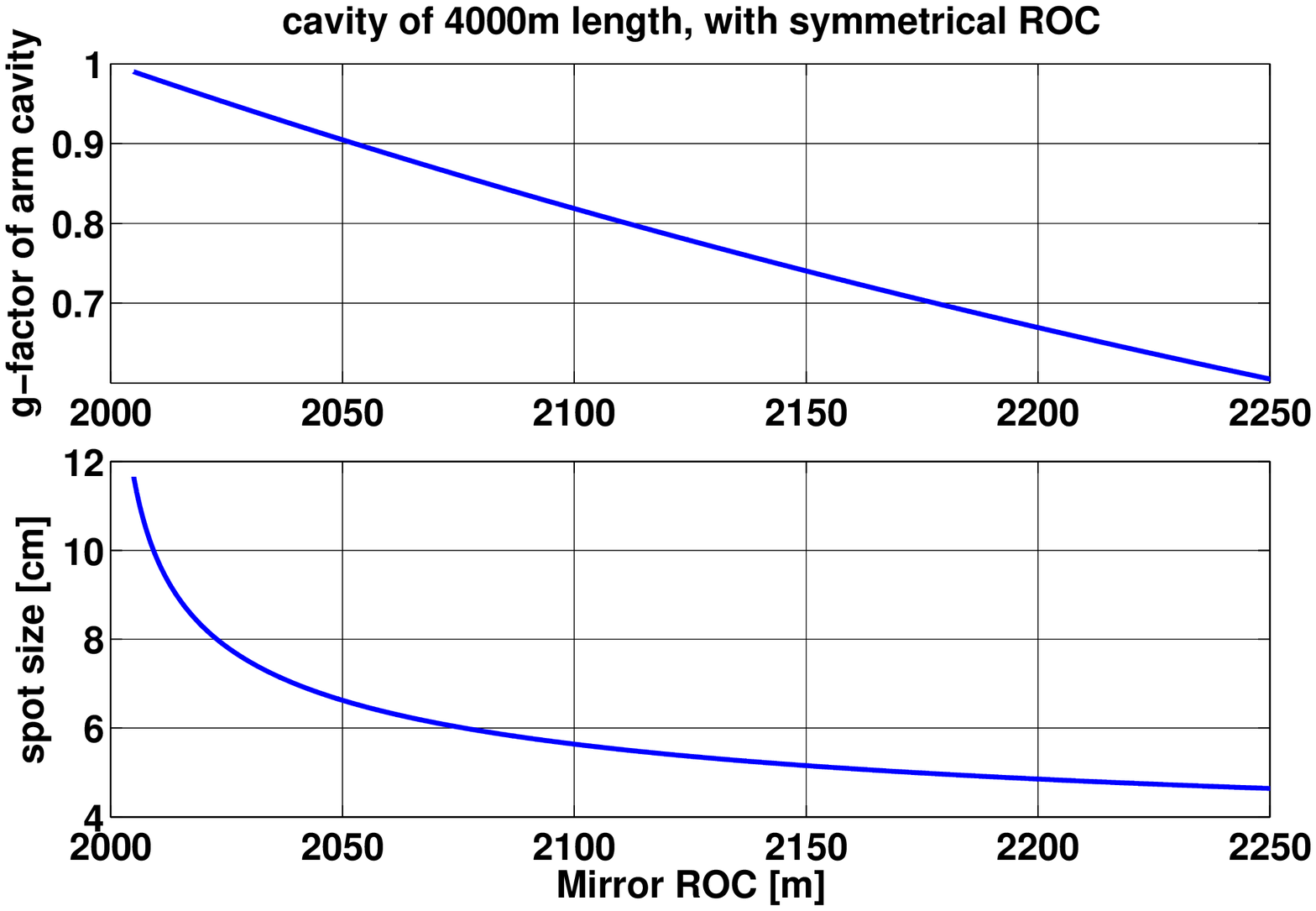}
\caption{LEFT: Quantum noise suppression factor  versus 
 losses along the path of the squeezed light from the generation to the detection. 
The differently coloured traces represent various initial squeezing levels. The plot illustrates 
the importance of a low loss implementation of squeezed light in GW detectors. Even
for a source with infinite squeezing level (red trace) it will be challenging to reduce the losses 
far enough to achieve a quantum noise suppression better than a factor of 3.  RIGHT:
Cavity g-factor and laser beam radius at the test masses of a 4km-long Fabry-Perot cavity
 versus the radius of curvature of the two cavity mirrors. Coating Brownian noise
decreases from right to left, while at the same time the cavity comes closer
and closer to its stability limit ($g \to 1$). }\label{fig:squeez-level}
\end{figure}

  \item{\textbf{Coating Brownian noise:}  The techniques under consideration for 
the reduction of the Brownian noise of the dielectric mirror coatings can be divided
into two strands: The first and more obvious class tries to directly reduce the 
coating noise by applying coating materials or doped materials with better mechanical properties
(see for instance \cite{Harry07}), making use of optimised (non-quaterwave)
coating layer thickness \cite{Agresti} or employing
micro-structure coatings (so-called wave-guide mirrors),
which can yield as high reflectivity as conventional
coatings but with significantly fewer coating layers \cite{Bunkowski06,
Friedrich11}. In addition the coating noise can be pushed
down by reducing the coating temperature \cite{clio_proof, Martin08}, which 
however would require significant hardware modifications and usually also demand
a change of the test mass and coating materials. The second 
class of techniques tries to reduce the effective coating noise level sensed
by the laser beams. The simplest way of doing so would be to increase
the beam size on the mirrors and therefore better averaging over the
thermal fluctuations. However, the maximum feasible beam size may be limited 
by the size of the vacuum tubes, by the stability of the cavities (g-factor
$\to$
1, see right plot of Figure \ref{fig:squeez-level}), by the commercially available
 maximal mirror substrate size or any combination of these
three issues. More challenging interferometric techniques capable of reducing 
the coating contribution include the application of non-TEM$_{00}$ laser beam profiles,
which yield the readout of a larger effective mirror surface area \cite{D’Ambrosio, 
Bondarescu, Mours, Chelkowski}. Furthermore, it has been
proposed to decrease the coating noise level by replacing the arm
cavity mirrors by anti-resonant cavities or etalons \cite{Khalili05,
Gurkovsky}. Again it has to be noted that the above
mentioned techniques are not exclusive, but can often be
combined. }

  \item{\textbf{Suspension Thermal noise:}  In principle the least disruptive way to reduce
suspension thermal noise, is to change the material, especially of
the last stage fibres, by one with better mechanical properties. 
It has also been shown that a further reduction of suspension thermal noise
can be achieved by improvements to the fibre profile, especially at the 
fibre necks \cite{Cumming}. More hardware-intensive improvements to suspension thermal
noise include macroscopic changes of the suspension dimensions and 
cooling the relevant suspension elements to cryogenic temperatures \cite{Puppo}.
 }
\begin{figure}[htbp]
\centering
\includegraphics[width=0.55\textwidth]{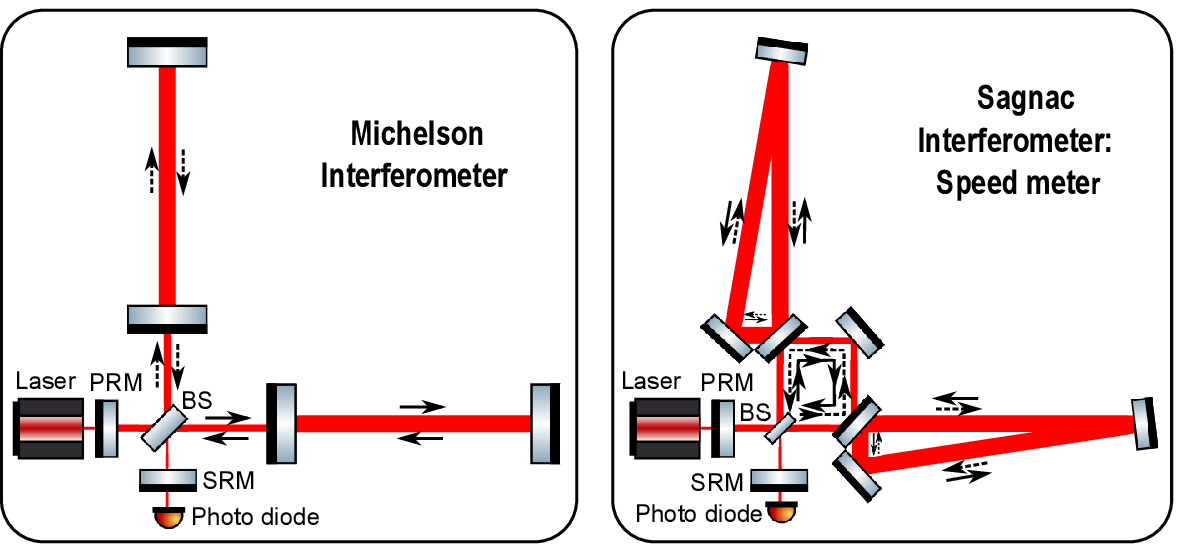}
\includegraphics[width=0.44\textwidth]{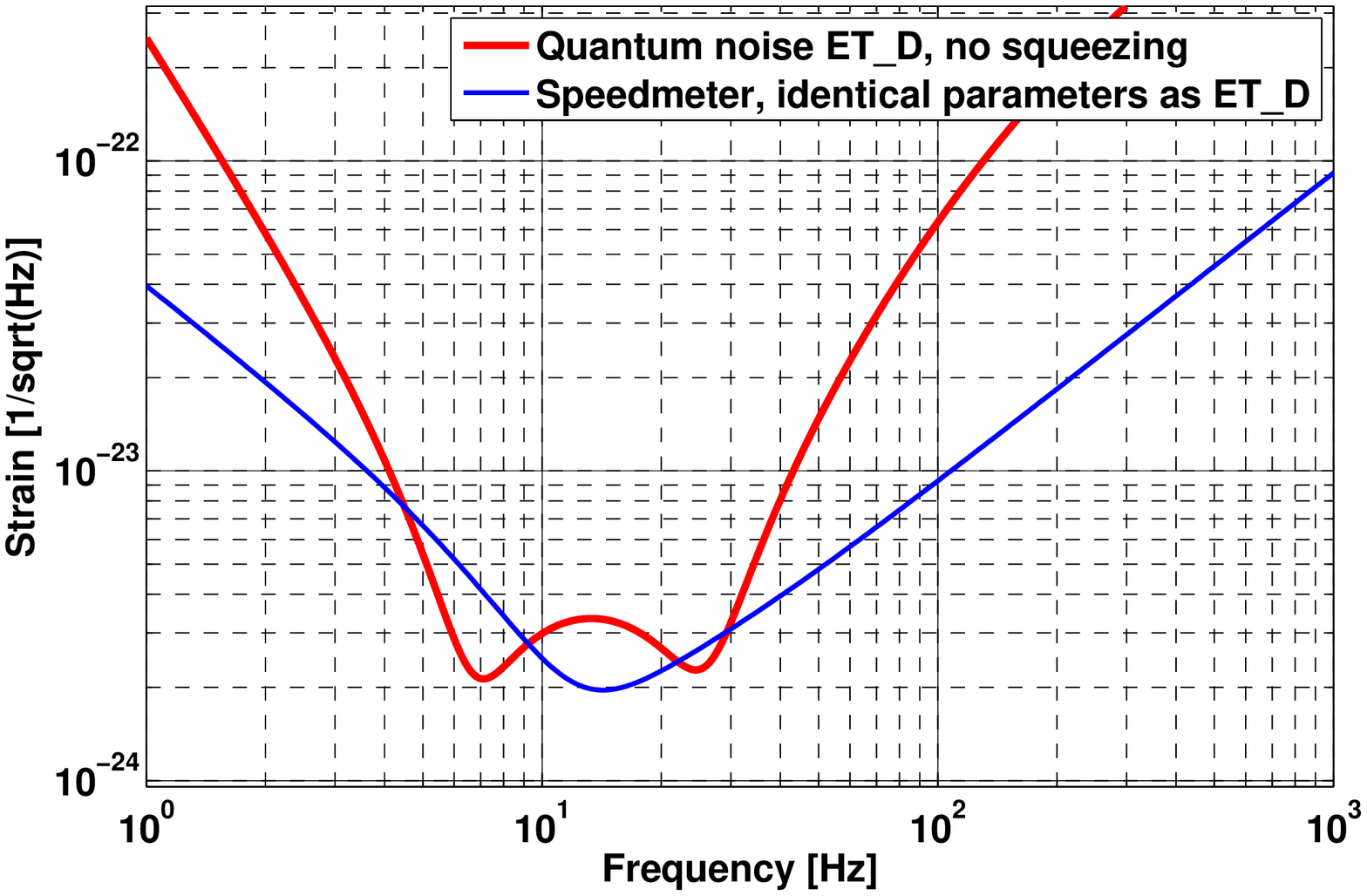}
\caption{LEFT: Simplified schematic of a Michelson interferometer with Fabry-Perot arm cavities
as well as power and signal recycling. CENTER: Simplified schematic of a Sagnac interferometer 
featuring arm cavities 
as well as power and signal recycling. RIGHT: Quantum noise limited sensitivity of Michelson 
interferometer with detuned signal recycling
versus a Sagnac speedmeter with similar parameters. As the speedmeter allows to suppress back
action noise (i.e. quantum radiation pressure noise) it provides better low frequency sensitivity 
and can therefore be designed to give a significantly 
larger detector bandwidth for roughly the same peak 
sensitivity.}\label{fig:speedmeter}
\end{figure}

  \item{\textbf{Seismic noise:}  The test mass displacement driven by direct 
coupling of seismic noise can be reduced by improving the seismic 
isolation systems. This can be achieved by either increasing the number of isolation 
stages in passive systems \cite{Ballardin} or by reducing sensor and control noise 
in active seismic
isolation systems \cite{Abbott04}. Due to the steep slope of seismic noise,
any major improvements of the seismic noise level rather requires a shift of the seismic 
noise wall towards lower frequencies than just improvement of the isolation system by a
small
factor \cite{Braccini10}. One special case of an active seismic isolation is the so-called
suspension point interferometer \cite{Aso}, which uses interferometric sensing techniques to 
stabilise the suspension points of several test masses with respect to each other. 
A completely different approach  to reduce the 
seismic noise contribution is not to minimise the coupling of the seismic from the ground
to the test mass, but to reduce the initial seismic excitation of the ground by
building the GW observatory on a seismically quiet location, for instance underground \cite
{Ohasi}.}
  \item{\textbf{Gravity gradient noise:} In contrast to
seismic coupling via the suspension system to the test
masses, which can be tackled by  better seismic isolation
systems, there is no way to shield the test masses from
acceleration caused  by seismically driven 
fluctuations of the gravitational potential. Therefore, the only two discussed
methods to reduce gravity gradient noise are building the GW observatory in a
location with low intrinsic seismic noise, for instance underground \cite{BekerGGN}
or the application of feedforward or subtraction techniques based on seismic sensor
 signals \cite{CellaGGN, HarmsGGN}}.  
  \item{\textbf{Residual gas pressure noise:}   The only feasible option to reduce the 
residual gas pressure noise is to improve the vacuum inside the beam tubes. }
\end{itemize}

From the above discussion it is clear that there are plenty
of options to improve the sensitivity of the second
generation instruments significantly beyond their initial
target sensitivity. Figure \ref{fig:ALIGO} shows the
Advanced LIGO sensitivity together with an orange area
which indicates a range of illustrative sensitivity limits of advanced
LIGO upgrades (see caption for exact description). This plot also
suggests a lack of a well
defined limit beyond which improvements of second generation cannot be pushed
any further. In the end the limits for any improvements are likely to be determined by
the point at which further upgrades of the second generation GW observatories 
will  cost more than it would to reach a similar sensitivity with 
less hardware effort in a new facility.

\section{The Einstein Telescope: A third generation gravitational wave observatory}
\label{sec:ET}

In 2008 work started on a Framework Programme 7 (FP7) funded
design study for a third generation gravitational wave
observatory, named the Einstein GW Telescope (ET), aiming for a 10 times increased  
sensitivity compared to second generation instruments. In addition one of the 
major goals of this work was to evaluate the possibility of pushing the observation band
down to frequencies as low as 1--2\,Hz. A detailed description of the 
completed design study \cite{ET-DS}  is beyond the scope of this
article, but a brief overview of the corner stones of the ET
design is given below.

\begin{table}[htbp]
\begin{center}
\begin{tabular}{l l l}
\hline 
\hline
Parameter & ET-HF   & ET-LF \\
\hline
Arm length & 10\,km & 10\,km \\
Interferometer type & FP-MI with DR & FP-MI with DR\\
Input power (after IMC) & 500\,W & 3\,W \\
Arm power & 3\,MW & 18\,kW\\
Temperature & 290\,K &  10\,K  \\
Mirror material & Fused silica & Silicon \\ 
Mirror diameter / thickness & 62\,cm / 30\,cm & min 45\,cm/ TBD \\
Mirror masses & 200\,kg & 211\,kg \\
Laser wavelength & 1064\,nm & 1550\,nm \\
SR-phase & tuned (0.0) & detuned (0.6)\\
SR transmittance & 10\,\% & 20\,\% \\
Quantum noise suppression &  freq. dep. squeez.& freq. dep. squeez. \\
Filter cavities & $1 \times 500\,$m  & $2 \times 10\,$km\\
Squeezing level  & 10 dB (effective) & 10 dB (effective) \\
Seismic isolation & SA, 8\,m tall & mod SA, 17\,m tall \\
Seismic (for $f>1$\,Hz) & $5\cdot 10^{-10}\,{\rm m}/f^2$ & $5\cdot 10^{-10}\,{\rm m}/f^2$  \\
\hline
\hline
\end{tabular}
\caption{Summary of the key parameters of the ET high and low-frequency
interferometers \cite{ET-D}. FP-MI with DR = Michelson with Fabry-Perot arm cavities and dual 
recycling, SA = super attenuator,  freq. dep. squeez. = 
squeezing with frequency dependent angle.\label{tab:summary}}
\end{center}
\end{table}

The ET observatory will be built in an underground location in order to suppress 
seismic noise  and associated gravity gradient noise, as well as to simplify potential gravity
noise subtraction schemes. The
observatory will have the overall shape of an equal-sided triangle of 10\,km
length, housing three GW detectors with an opening angle of each 60 degrees
and therefore allowing to fully reconstruct the polarisation of the GW source as 
well as providing redundancy \cite{trimi, triangle}.

Initial design efforts for ET focussed on using one
interferometer to cover the full frequency range from
1--10000\,Hz \cite{ET-B}. However, analysis of competing noise sources
and their corresponding design requirements
\cite{FreiseGRG}, revealed the difficulty of designing one
interferometer that could be extremely sensitive over the
full frequency range. For example, achieving good low 
frequency sensitivity requires cryogenic test masses
to minimise the various thermal noise contributions, but at
the same time one would need to use optical power in the
megawatt range to obtain good high frequency sensitivity.
Residual absorption in the test masses and their coatings
would then cause considerable amounts of heat to be deposited in
the test masses, and this would need to be extracted via the mirror
suspension fibres. It turns out this would set impractical requirements on
the suspension fibres and their associated thermal noise. 
Therefore, the ET baseline design adopted a `xylophone'
strategy \cite{Shoemaker01, conforto03, DeSalvo04} and each
of the three detectors within the triangle will consist of
two individual interferometers, one optimised for the 
low-frequency range and the other for the high-frequency
range \cite{ET-C}.

\begin{figure}[htbp]
\centering
\includegraphics[width=0.54\textwidth]{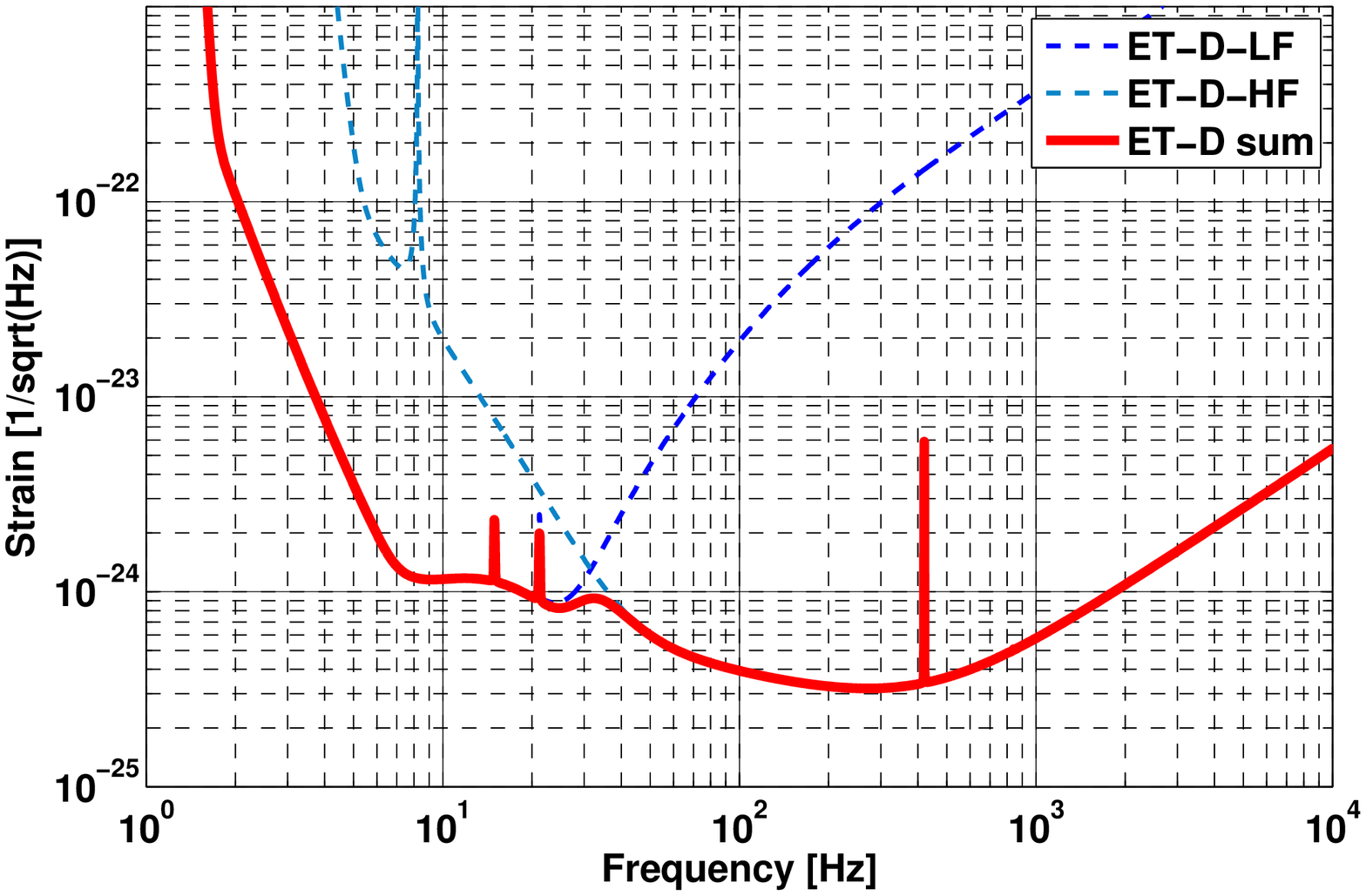}
\includegraphics[width=0.45\textwidth]{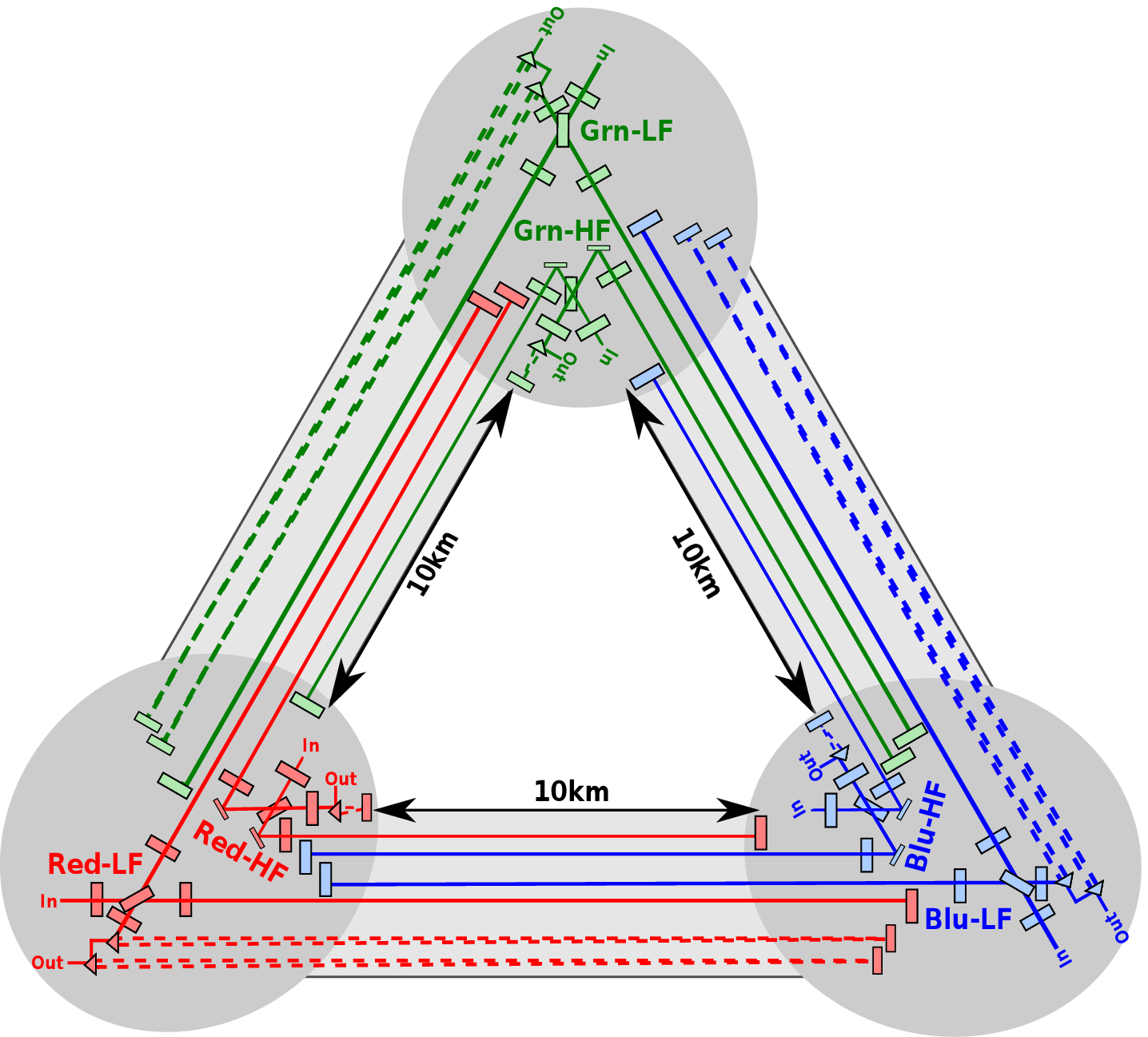}
\caption{LEFT: Sensitivity curves for the ET low-frequency (LF) and 
high-frequency (HF) interferometers together with the total sensitivity 
curve \cite{ET-D}. RIGHT: Simplified layout of the core interferometers
of the ET observatory. The Observatory is made out of three xylophone 
detectors (red, blue green) with 10\,km arm length, each consists 
of two interferometers, one covering the low-frequency range and one 
the high-frequency range. The solid lines indicate laser beams of the core
interferometers, while the dashed lines indicate filter cavities for the 
 injection of frequency dependent squeezed light.}\label{fig:ET}
\end{figure}

Table \ref{tab:summary} gives an overview of the key design
parameters of of the ET-baseline configuration  \cite{ET-D}.
While the high-frequency detector features technologies
similar to what an upgraded second generation detector might
look like, the ET low-frequency detector features a
low-power (only 18\,kW of circulating power), low-temperature 
(10\,K) design. Going from room temperature
operation to cryogenic temperatures
requires a change of the test mass material \cite{Wider00} from fused silica to
silicon \cite{Rowan03}, which in turn demands a change of laser
wavelength from 1064\,nm to $\approx$1550\,nm. In addition the ET low-frequency
interferometers employ two 10\,km long low-loss, filter-cavities for
the generation of frequency dependent squeezed light  
\cite{ET-0104A-10}. The left plot of Figure \ref{fig:ET} shows the
resulting
sensitivities of the ET low-frequency and high-frequency 
interferometers, together with their combined sensitivity.  A simplified 
layout of the full ET observatory is shown in the right plot of   Figure
\ref{fig:ET}.

\section{Summary and Outlook}
\label{sec:summary}
As we have seen, the baseline designs of the second generation gravitational
wave detectors will not exhaust the sensitivity limits of their
facilities. A multitude of techniques has been suggested
to upgrade  the second generation instruments.
Currently efforts are on the way for down selection of these
technologies and the preparation of design proposal for
second generation upgrades.

With the completion of the ET design study, a baseline
design for a third generation gravitational wave detector has been presented. Over
the next few years this design will be further refined. 
The key technologies required to build future gravitational
wave detectors have been identified over the past few years and 
current and future research efforts will reveal which of
these technologies will provide the highest robustness and the 
best sensitivity gain.

\ack{The author thanks H.~L\"uck, D.~Shoemaker and K.~Strain
for valuable comments on this manuscript. In addition the author would like to 
thank all members of the LIGO Scientific 
Collaboration and the ET-Science Team (Grant Agreement 211743) for their 
contributions to the presented work.  The 
author is grateful for the support from the Science and Technologies Facility 
Council (STFC). }

\end{document}